# Investigation of Gaze Patterns in Multi View Laparoscopic Surgery

Navaneeth Kamballur Kottayil[1], Rositsa Bogdanova[2], Irene Cheng[1], Anup Basu[1] and Bin Zheng[2]

*Abstract*— Laparoscopic Surgery (LS) is a modern surgical technique whereby the surgery is performed through an incision with tools and camera as opposed to conventional open surgery. This promises minimal recovery times and less hemorrhaging. Multi view LS is the latest development in the field, where the system uses multiple cameras to give the surgeon more information about the surgical site, potentially making the surgery easier. In this publication, we study the gaze patterns of a high performing subject in a multi-view LS environment and compare it with that of a novice to detect the differences between the gaze behavior. This was done by conducting a user study with 20 university students with varying levels of expertise in Multi-view LS. The subjects performed an laparoscopic task in simulation with three cameras (front/top/side). The subjects were then separated as high and low performers depending on the performance times and their data was analyzed. Our results show statistically significant differences between the two behaviors. This opens up new areas from of training novices to Multi-view LS to making smart displays that guide your shows the optimum view depending on the situation.

## I. INTRODUCTION

Laparoscopic Surgery (LS) is becoming the standard procedure in surgery. It uses small incisions created in the patients body to insert surgical tools and camera. The operation is performed with the surgeon guiding the surgical tool with the camera guiding his movements. LS offers a lot of benefits to the patients including reduced recovery times, less risk of hemorrhaging etc. However, for the surgeons using the system, it is difficult to perceive the surgical site in 3-dimensional (3D) fashion and coordinate their eyes and hands, due to the loss of depth perception, indirect image, mirrored hand movements, and eye- hand misorientation using a single camera[1]. The limitation of visual perception in LS increases cognitive and physical stress of the surgeons and trainees and is a leading cause of inaccurate judgment and estimation. This leads to significantly longer times in training and performing LS [2].

Some recent studies have found that different camera arrangements affect perceptual-motor performance in laparoscopic surgery [3] - [5]. The use of multiple cameras as a tool for restoring the three dimensionality is optimistic and can easily resemble the different vantage points accessibility of open surgery. In the current study we investigate the behavior of subjects when presented with multiple viewing perspectives in surgical simulation. We compare the eye behavior of the high and the low performers when attempting to perceive the depth cues presented with a multiple view setting.

*This work was not supported by any organization
[1]Dept of Computing Science, University of Alberta
[2]Department of Surgery, University of Alberta

## II. MOTIVATION

It was shown that a multiple view arrangement can be superior to the use of a single camera [3] - [5]. However, this also increases the cognitive load on the user. From studies in aviation displays, we have the conclusion that mentally integrating information across multiple displays is challenging and draws additional attention demands from the user [6] - [8]. In the study of DeLucia on effects of camera arrangement on perceptual-motor performance in LS, multiple camera views provided more information about 3D space but imposed more attention demands compared with a single-camera view. In their study participants were presented with multiple-camera views of a surgical simulation environment. Participants did not look at all views equally often and may not have necessarily mentally integrated the views to reconstruct 3D space [4]. It was also suggested that surgeons (elite performers) use different information or integrate multiple sources of information differently than novices. This leads us to believe that with training, humans might learn a specific 'gaze behavior' that integrates the 3D information more efficiently. We seek to discover this behavior that separates the experts from the novices. The results can help in design of better displays in multi-view environment, more intelligent camera placement and better training programs for novices [9].

## III. METHOD

In the study we conducted, we compared the gaze behaviors of human operators while performing simple surgical task in a multiple camera view condition.

### A. Subjects and experimental environment

Twenty university students with varying levels of laparoscopic surgical training participated in the study (13 males, 7 females; mean age, 28). The sample size was calculated based on outcomes from previous research found in the literature. For example, we used DeLucias study (2011) to adjust our sample size. In DeLucias study, 12 subjects were included in testing how the number and type of camera views affect manual manipulation. They recorded a main effect of viewing condition ($F(4, 44)= 23.79$, $p < 0.001$, $w2= 45.58\%$). Tukeys HSD analyses showed that mean task completion time was significantly faster for the direct view compared with the front and side views, and the side view resulted in the slowest completion time among the different views ($p < 0.05$). The larger group size in the current study was expected to have significant power to show significant effects. Ethics approval was obtained from the Health Research Ethics Board of the University of Alberta before the recruitment of human subjects. Written consent was obtained from each participant prior to entering the study.

The experimental setup included three main components, (1) a 2D monitor (LG-24MA31D, LG Electronics, Seoul, South Korea) which displays images captured by the surgical cameras (2) Training box with three camera for front, top and side views and 3) Tobii X2 60Hz eye-tracker placed under the monitor to unobtrusively record the subjects eye motions.

Subjects performed a surgical simulation task with the camera placed at two different angles (front camera at 30-, and top camera at 90- degree angle to the plane of the target), with a third camera placed on the side of the training box.

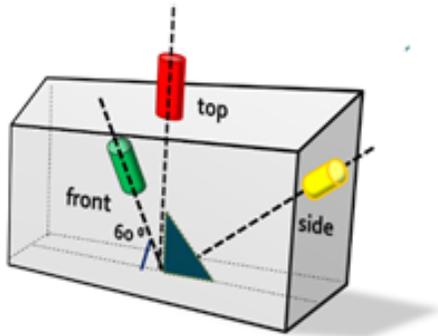

Fig. 1. 2D camera positions in training box.

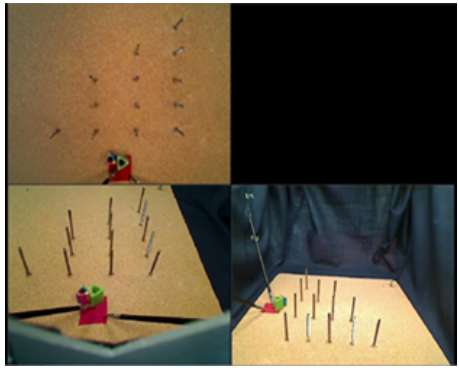

Fig. 2. Three views.

### B. Tasks

Subjects were asked to move graspers and transfer objects between surgical targets (pegs) in different depth planes (Figure 3). Subjects were required to perform the task as quickly and accurately as possible without dropping the objects. During the entire trial subjects eye movements were remotely recorded by an eye-tracking system

## IV. ANALYSIS

Careful analysis on eye-gaze behaviors of the higher and the lower performance trials was conducted by analyzing the eye tracking data from Tobii Studio. We also built a proof of concept system for analyzing the object distances and velocities at each subtask. To do this, we used Kanade-Lucas-Tomasi (KLT), feature-tracking algorithm [10] for tracking the objects. The algorithm can be used to track a set of feature points using optical flow estimation. We used an opencv implementation of the same.

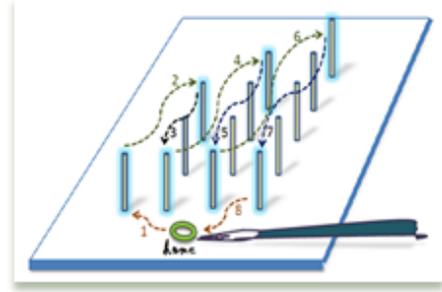

Fig. 3. Task that the subjects were asked to complete.

Our implementation re-initializes the tracking every 10 frames or every time there is less than 2 feature points that are tracked. The feature points we tracked are found using color thresholding and feature detection algorithm proposed in [11] that detects strong corners of the image from the video frame. A screen grab from the actual system showing the object detection and tracking is shown in Fig 4.

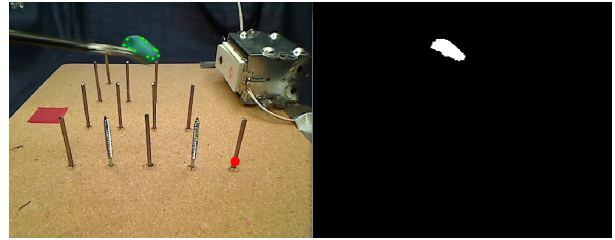

Fig. 4. Object detection and tracking; the green spots denote the tracked points, the red spot denotes the target point. The figure on the right shows a segmented view of the pixels being tracked.

*Detection of task completion and next target:* In the series of tasks we asked the user to perform, the target kept changing after the user performed the task. We asked the experimenter to mark all the targets of the task that the user has to perform. The system would cycle through these points to detect task completion.

### A. Identification of experts using time performance

We used time of task completion metric to measure performance. All 20 subjects performed the study task in the multiple viewing condition. The average completion time was 4.89 min (min: 2.2; max: 8.37). We selected the fastest 25 % as high performers and the slowest 25% as low performers for conducting further analysis.

*Statistical tests*: By comparison of the low and high performers eye behavior we were able to reveal how human operator collect visual information to rebuild 3D vision in Image-guided surgery. The eye behavior measurements

(percentage of view used and frequency of gaze shift) were subject to 2 (groups; high performers versus low performers) x 2 (views; top versus front) two-way ANOVA model for analysis of variance. To define further difference between the high and the low performance groups, data was subjected to a independent samples t-test. SPSS (SPSS Inc. Chicago) was used to perform statistical analysis and p ¡ 0.05 was considered statistically significant.

*B. Features used*

From the eye tracker and the feature tracking, we analyzed the following data:

1) Gaze location: Initial analysis of subjects gaze location over the three camera views subjects spent sufficient time (48%) on the top view as well as the front view (50%), but not on the side view, which was used at only 1%. Additionally, we compared the number of visits between the 3 views. Pair comparison revealed significant difference between the top and the side view (p < 0.001); front and side view (p < 0.001), but not between top and front view (p = 0.971). For this reason, we excluded the side view data in our further comparisons.
2) Frequency of gaze shift between the different views: This looked at the amount of time there was a shift of gaze from one view to other. The statistical test results showed that there is significant main effect between the high and low performers : $F(3,16) = 8.96$; $p = 0.009$; $\eta^2 = 0.359$. The graphical representation of the same is show in 5

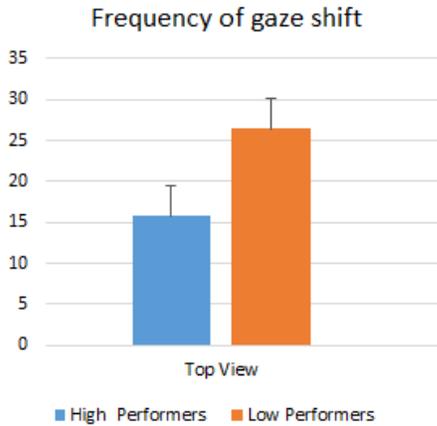

Fig. 5. Number of time Gaze of the user shifted between top and front view by the high and low performers.

3) Percentage of view used (eye behavior measurements): This looked at the amount of time one particular view was utilized during the course of the experiment by a user. We consider percentage values here to remove the effect of varying times among users. Results revealed for the group as a whole, significant main effect for views: $F(3,16) = 27.71$; $p = < 0.001$; $\eta^2 = 0.634$; but not significant main effect for groups: $F(3,16) = 0.02$; $p = 0.89$; $\eta^2 = 0.001$; and significant interaction effects: $F(3,16) = 6.56$; $p = 0.02$; $\eta^2 = 0.29$. Results are in Fig 6

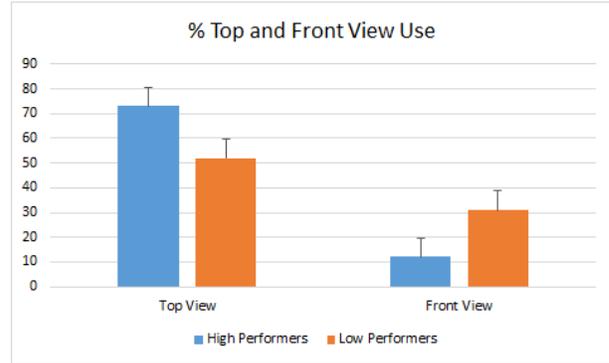

Fig. 6. Percentage of Top and Front view used by the high and the low performers.

4) Front/Top ratio is obtained by dividing the percentage of gaze on the front view to the percentage of gaze on the top view. The Front/Top ratio was significantly different between high performers (M = 0.19 ± 0.15) and low performers (M = 1.02 ± 1.12) groups: $t(8) = -1.64$; $p = 0.034$. Graphical results are shown in Fig 7

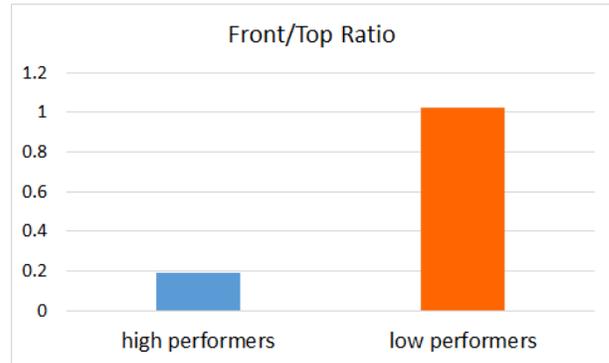

Fig. 7. Front/Top ratio for high and low group performers.

5) On screen distance ($\delta$) defined as:

$$\delta = min(dtop, dfront) \quad (1)$$

where dtop, dfront are the on screen distance of the tracked object to the target in the top and front views respectively. A plot of on screen distance vs amount of time front and top views were used for the experts and novices for the use of top view and the front view are shown below for distances up to 200 pixels (Sample plots of high and low performers in Fig 8, 9).

## V. CONCLUSION

Based on the results of our study, we can underline a few key points of importance. We confirm the theory that additional views contain important information for subjects to perceive the depth of the surgical site. This was proven with the finding that 48% of the time, subjects used the top view for guiding their performance along with the conventional (front) view used at 51%.

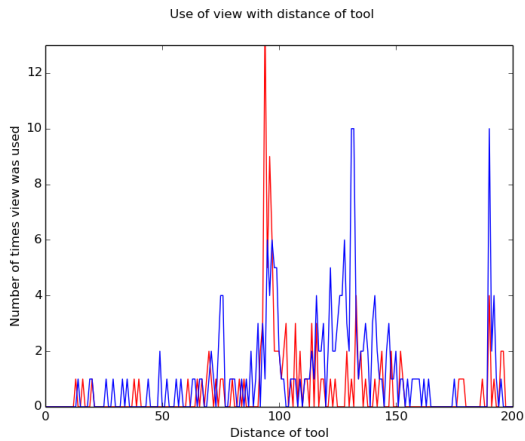

Fig. 8. High performer. Here red plots are the cases where the top view was used. The blue are the cases where the front view was used.

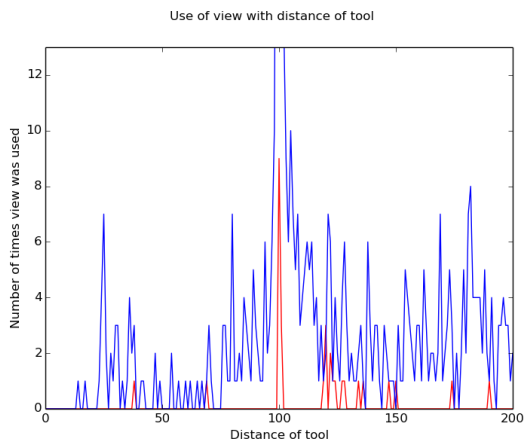

Fig. 9. Low performer. Here red plots are the cases where the top view was used. The blue are the cases where the front view was used.

Our results revealed that the high performance group used the top view more than 70% of the trial time compared to the front view, which they referred to only 12% of the time. This was particularly so when the object distance was closer to the target. This behavior was confirmed by the detailed analysis of the camera use at specific distances from the target shown in Fig 8,9. Other time was spent on view transition, saccades, blinking, and side view visits.

We found that the members of the high performance group shifted their gaze between the top and the front view less frequently compared to their low performance counterparts. This finding suggests that the experts performers exhibited more concentrated to a particular view eye behavior whereas the novices were frequently shifting their visual attention between views

High performers used both views in more balanced way throughout the trial compared to their low performance counterparts which focused mainly on one view. We conclude these from the Front/Top ratio for the high and low performers.

In conclusion, the current study showed that human operators utilize information from different visual sources, when available, for reconstructing the three-dimensionality of a surgical scene without impairment of performance. We additionally revealed eye behavioral evidence to support the notion that expert and novice performers use the visual information from the multiple sources differently. As was suggested in previous research on perceptual-motor performance in the LS [3], top surgeons (elite performers) might use different information or integrate multiple sources of information differently than novices. It is also believed that, with practice, trainees could learn to use the multiple views to improve performance beyond what was achieved with a single view. Thorough understanding of the operators behaviors is vital to direct further research on the feasibility of multiple views usage in the OR and for surgical training. The knowledge from this study can be used for creation of smart display interfaces where trainees gaze can be guided towards an expert-like behavior. The benefit of such a remote training tool (smart display system) is significant.